# ESKNet: An enhanced adaptive selection kernel convolution for ultrasound breast tumors segmentation

Gongping Chen[a,b,1], Lu Zhou[a], Jianxun Zhang[a], Xiaotao Yin[c], Liang Cui[d], and Yu Dai[a]

*E-mail:* cgp110@ mail.nankai.edu.cn; zhoulu@nankai.edu.cn; zhangjx@nankai.edu.cn; yxtfwy@163.com; m15225042110@163.com; daiyu@nankai.edu.cn

[a] *The College of Artificial Intelligence, Nankai University, Tianjin, China*

[b] *The School of Biomedical Engineering and Technology, Tianjin Medical University, Tianjin, China*

[c] *The Department of Urology, Fourth Medical Center of PLA General Hospital, Beijing, China*

[d] *The Department of Urology, Civil Aviation General Hospital, Beijing, China*

**Abstract:** Breast cancer has become one of the most dreaded diseases that can threaten the life of any woman. Accurate target lesion segmentation is essential for early clinical intervention and postoperative follow-up. Recently, many convolutional neural networks (CNNs) for segmenting breast tumors from ultrasound images have been presented. However, the complex ultrasound pattern and the variable tumor shape and size bring challenges to the accurate segmentation of the breast lesion. Motivated by the selective kernel convolution, we introduce an enhanced selective kernel convolution for breast tumor segmentation, which integrates multiple feature map region representations and adaptively recalibrates the weights of these feature map regions from the channel and spatial dimensions. This region recalibration strategy enables the network to focus more on high-contributing region features and mitigate the perturbation of less useful regions. Finally, the enhanced selective kernel convolution is integrated into U-net with deep supervision constraints to adaptively capture the robust representation of breast tumors. Using three public breast ultrasound datasets, we conducted extensive experiments with many state-of-the-art deep learning segmentation methods. In the segmentation of the first ultrasound dataset (BUSI), the values of Jaccard, Precision, Recall, Specificity and Dice are 70.20%, 79.57%, 82.41%, 97.47% and 78.71%, respectively. The values of Jaccard, Precision, Recall, Specificity and Dice for our method on the second ultrasound dataset (Dataset B) are 71.65%, 81.01%, 82.66%, 99.01% and 79.92%, respectively. For the segmentation of external ultrasound dataset (STU), the mean values of Jaccard, Precision, Recall, Specificity and Dice are 75.14%, 84.73%, 89.25%, 97.53% and 84.76%, respectively. The experimental results fully demonstrate the superior performance of our method for segmenting breast ultrasound images. The source code is available on the following website: https://github.com/CGPxy/ESKNet.

## 1. Introduction

Breast cancer is as one of the most popular malignant diseases in the women, which seriously endangered their health and even their lives (Chen et al., 2023a). Regular early screening is critical to develop the medical programme and reduce the mortality rate due to the strong concealment and many inducements of breast cancer (Xian et al., 2018). Currently, ultrasound imaging has become one of the most common technical means for clinical screening of breast cancer due to its non-invasive, cheap and fast superiority (Ilesanmi et al., 2021). Regrettably, accurately and quickly annotating the lesion regions is a challenge even for experienced radiologists due to the complexity of the ultrasound image, as shown in Fig. 1(a). To overcome this problem, various computer-aided diagnosis systems (CAD) have been established to assist doctors in interpreting breast ultrasound images (Huang et al., 2023a; Li et al., 2022; Xu et al., 2019). It is well known that medical image segmentation can help to locate and evaluate pathological regions (Chen et al., 2021). Therefore, medical image segmentation is one of the indispensable steps in a CAD system (Huang et al., 2023b, 2023c).

---

[1] Corresponding author

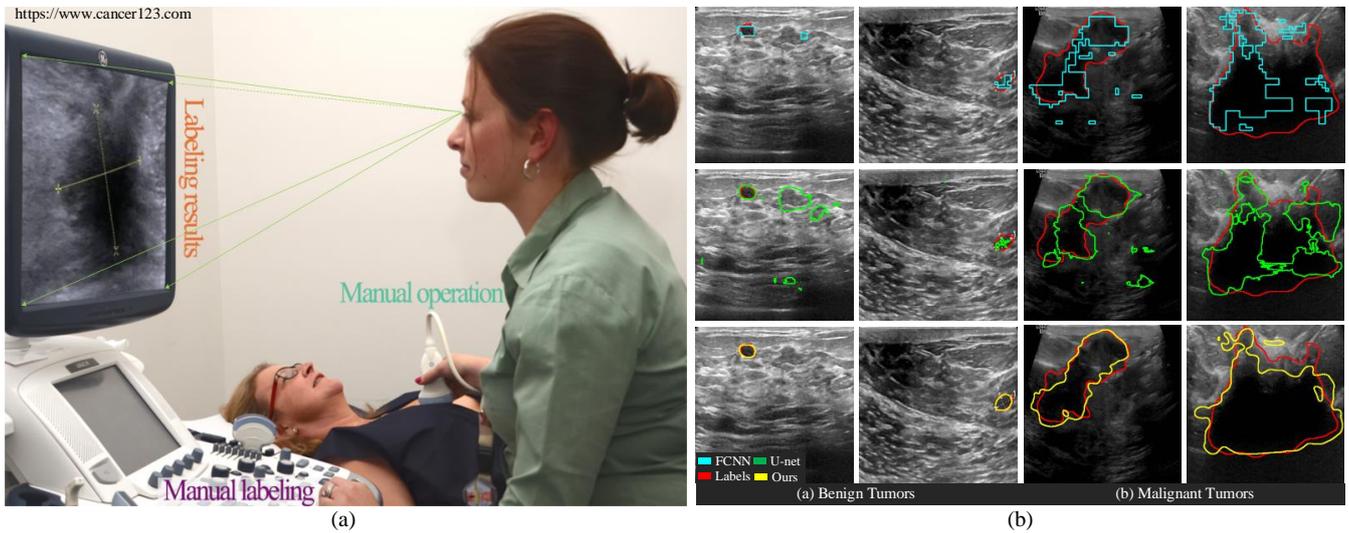

Fig. 1. The segmentation results of FCNN, U-net, and our method from top to bottom respectively. The red curve is the outline of breast tumors. As we can see, variable tumor morphology, blurred boundaries, heterogeneous structures, and similar surrounding tissues seriously affect the segmentation precision of lesion tissue.

Achieving accurate segmentation of breast tumor regions in ultrasound images has always been a subject of extensive research (Abdelrahman et al., 2021). Previously, manual prior methods were often used to fit the contours of breast tumors. Xue et al. pointed out that the limited representation ability of artificial prior can easy to cause the misrecognition of breast lesions in complex ultrasound images (Xue et al., 2021). In addition, the method based on manual prior requires a lot of time and manpower. Recently, many representative CNNs were applied to medical image segmentation successfully and extensively (Fu et al., 2022). Among many excellent segmentation methods, FCN and U-net (Ronneberger et al., 2015) are two unique representative works and have been applied in medical image segmentation extensively. Almajalid et al. used U-net for the first time to accomplish the task of breast ultrasound image segmentation (Almajalid et al., 2018). Subsequently, Yap et al. comprehensively analyzed the segmentation performance of three benchmark networks: U-net, FCN-AlexNet, and patch-based LeNet in coping with breast lesions (Yap et al., 2018). Similarly, Mishra et al. used FCN to design a deep supervision network for ultrasound image segmentation (Deepak et al., 2018). However, breast lesion segmentation is an extremely challenging job due to the complex ultrasound patterns and similar intensity distribution (Ning et al., 2021). Therefore, it is difficult to acquire satisfactory segmentation results by simply applying existing frameworks (such as U-net, FCN, etc.) on breast ultrasound images, as shown in Fig. 1(b).

For breast ultrasound, the main interfering factors that impede the accurate segmentation of breast lesions are the following: 1) similar intensity distribution and blurred boundary, especially in malignant lesions; 2) significant morphology and position variation of breast tumors (Chen et al., 2022a). If we want to obtain the precise segmentation results from the ultrasound images, the segmentation network design should not only be able to adapt to breast tumors with different scales, but also need to improve the focus on the lesion regions. The benefits of attention mechanisms and multi-scale convolution have been demonstrated in many low-level tasks (Elmoufidi, 2022; Tomar et al., 2022). The attention strategy can help the network to extract useful information that better characterizes the objective while reducing the introduction of useless information. Multi-scale convolution improves the representation ability of objects by using different convolution kernels to capture interesting target features from different scales of the receiving domain (Joshua et al., 2020). The CNN architecture based on attention mechanism and multi-scale convolution for breast ultrasound image segmentation task is extensively employed (Abdelrahman et al., 2021). For example, Yan et al. used the hybrid dilation convolution based on the attention U-net for proposing an novel attention-enhanced U-net (AE U-net) to automatically segment breast tumors in ultrasound images (Yan et al., 2022). For capturing objective features in the varying receptive fields, Zhuang et al. introduced dilated convolution and residual learning in the Att U-net (Zhuang et al., 2019). However, the use of dilated convolutions on deeper convolutional layers cannot capture sufficient contextual information (Xue et al., 2021). To better capture the multi-scale information of the breast tumors, Punn et al. (Punn and Agarwal, 2022) replaced the convolution blocks of Att U-net with the residual blocks constructed by the inception convolution layer. Abraham et al. (Abraham and Khan, 2019) constructed a new U-shaped network (MADU-net) to segment breast ultrasound images by introducing a multi-scale image input pyramid and deep supervision mechanism into the Att U-net. Multi-image inputs can provide more fine-grained feature maps, but introducing too many low-level feature maps will affect the characterize high-level semantic features and reduce the representational capability of segmentation networks. Although these methods improve the segmentation accuracy of breast tumors to varying degrees, they still have two obvious limitations: (i) The multi-scale information is more dependent on the artificially set convolution kernel size and cannot adaptively capture the multi-scale information of breast tumors (Li et al., 2019). (ii) They tend to use a single attention mechanism to calibrate objective features. Recently, Li et al. designed a selective kernel convolution (SK) to adaptively select useful feature information under different receptive fields, as shown in Fig. 2(a) (Li et al., 2019). Although the segmentation performance of the network was improved by introducing the SK block, the strategy has two obvious limitations: (i) ignoring the calibration of spatial dimension features, and (ii) reducing the correlation of features in the module.

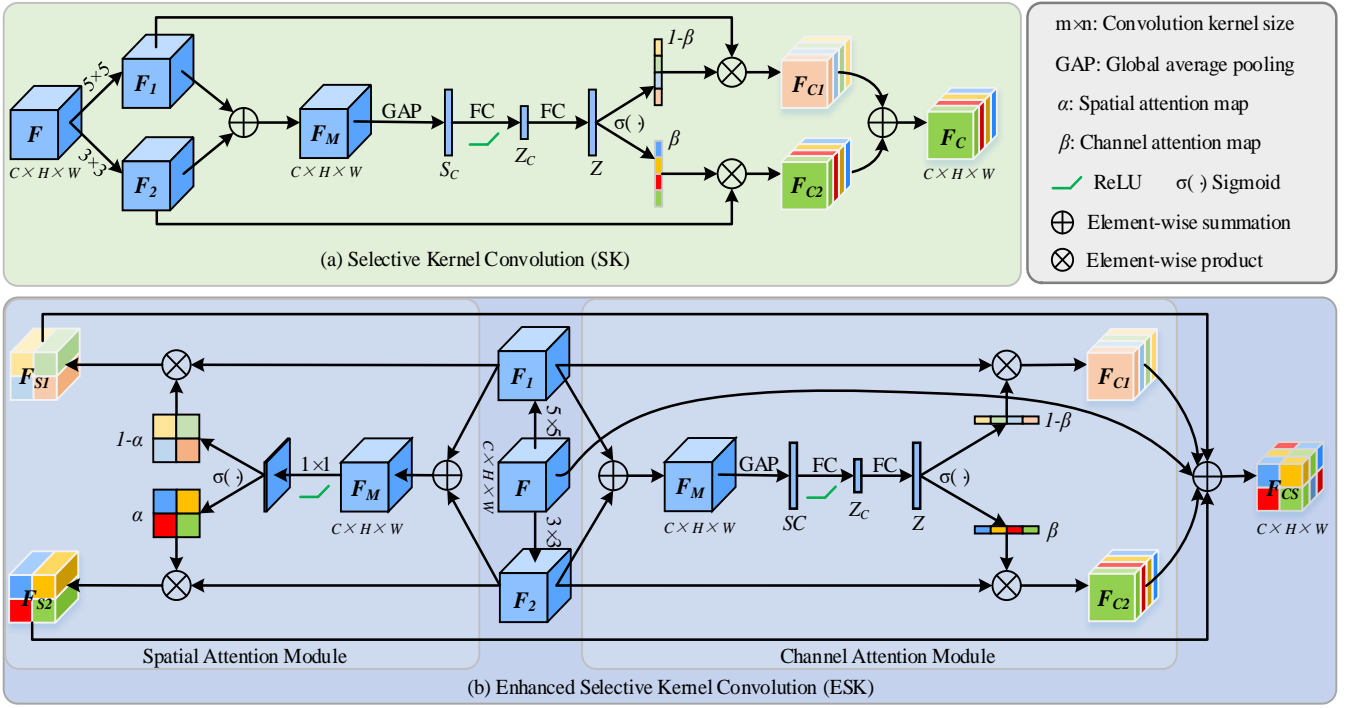

Fig. 2. The illustration of the enhanced selective kernel convolution (ESK, see Section 3.1) and the selective kernel convolution (SK).

To overcome above limitation, we first introduce spatial attention into the selective kernel convolution module to calibrate spatial dimension features, as shown in Fig. 2(b). Then, residual learning is added to the selective kernel convolution module to strengthen the relevance of long-distance features as shown in Fig. 2(b). Finally, we use the enhanced selective kernel convolution module (ESK) to construct a novel deep supervised U-net (Named as ESKNet, as shown in Fig. 3) to segment breast lesions adaptively within the ultrasound image. In general, the approach proposed in this paper can be summarized with the following characteristics:

- First, an enhanced selective kernel convolution is designed, which not only adaptively selects features under different scale receptive fields from the channel and spatial dimensions but also further strengthens the relevance of remote feature information.
- Second, a novel deep supervision U-net integrating the enhanced selective kernel convolution module is developed to segment breast lesions within the ultrasound image. The network can improve breast lesions segmentation accuracy by learning lesion-specific characterizations from breast ultrasound images.
- Moreover, we conducted extensive experiments with many state-of-the-art deep learning segmentation methods on the available datasets. The experimental results fully demonstrate the superior performance of our method for segmenting breast ultrasound images.

Table 1 The summary of literatures in "Related works".

| | Method | Year | Medical dataset | Main contributing components or operations |
|---|---|---|---|---|
| **CNNs for breast ultrasound segmentation** | ConvEDNet (Lei et al., 2018) | 2018 | Private Dataset | Transfer learning, Deep boundary supervision |
| | GG-Net (Xue et al., 2021) | 2021 | BUSI + Private Dataset | Boundary detection, Global guidance block (Channel and Spatial) |
| | C-Net (Chen et al., 2022b) | 2022 | BUSI | Residual-learning, Global guidance (Spatial attention) |
| | STAN (Shareef et al., 2020) | 2020 | Dataset B + BUSIS | Multi-scale |
| | SKU-net (Byra et al., 2020) | 2020 | Dataset B + BUSI + OASBUD | Multi-scale, Parallel channel attention |
| | SBA (Luo et al., 2022) | 2022 | Private Dataset (Dataset C) | Two parallel feature extraction networks, Channel attention |
| | AMS-PAN (Lyu et al., 2023) | 2023 | BUSI + OASBUD | Multi-scale, Channel attention, Spatial attention |
| | IU-net MALF (Tong et al., 2021) | 2021 | Private Dataset | Residual-learning, Spatial attention, Mixed attention loss |
| | RDAU-net (Zhuang et al., 2019) | 2019 | Dataset B + STU | Spatial attention, Residual-learning, Hybrid dilated-convolution |
| | CAD (Moon et al., 2020) | 2020 | BUSI + Private Dataset | Transfer learning, Image fusion |
| | Wang et al. (Wang et al., 2019) | 2019 | Private Dataset | Deep-supervision, Threshold loss |
| **Attention mechanism** | Att U-net (Oktay et al., 2018) | 2017 | Non-medical Dataset | Spatial attention |
| | SE block (Hu et al., 2020) | 2020 | Non-medical Dataset | Channel attention |
| | scSE block (Roy et al., 2018) | 2018 | MRI brian + Visceral dataset | Channel attention, Spatial attention |
| | SANet (Zhong et al., 2020) | 2020 | Non-medical Dataset | Residual-learning, Channel attention |
| | UNETR (Hatamizadeh et al., 2022) | 2022 | BTCV Dataset + MSD Dataset | Transformer attention encoder |
| | Swin-Unet (Cao et al., 2022) | 2022 | Synapse Dataset + ACDC Dataset | Swin Transformer attention |
| | MSSA-Net (Xu et al., 2021) | 2021 | Dataset B + BUSIS + Dataset C | Multi-scale self-attention |
| | Focal U-Net (Zhao et al., 2022) | 2022 | Private Dataset | Focal self-attention |
| | SK (Li et al., 2019) | 2019 | Non-medical Dataset | Multi-scale, Parallel channel attention |
| | Ours | 2023 | Dataset B +BUSI +STU | Multi-scale, Channel attention, Spatial attention, Parallel channel attention, Residual-learning, |

## 2. Related works

### 2.1 CNNs for breast ultrasound segmentation

Many advanced CNNs have obtained better results than traditional segmentation methods in breast lesion segmentation (Houssein et al., 2021; Xian et al., 2018). Lei et al. improved the segmentation performance of the encoder-decoder network for whole breast ultrasound images through introducing a boundary regularization strategy (Lei et al., 2018). To alleviate the problem of low correlation of distant features, Xue et al. used a boundary detection module and a global guidance block to develop a global guidance network to achieve automatic segmentation of breast lesions (Xue et al., 2021). Similarly, Chen et al. constructed a cascading network architecture to segment breast ultrasound images by designing a bidirectional attention network, which could constrain the segmentation results from a more global perspective (Chen et al., 2022a). During the segmentation of breast lesions, the introduction of boundary constraints can refine the prediction results of the network to varying degrees, but it is still challenging to acquire the precise boundaries from heavily cascaded or shadow-occluded regions. To cope with the challenge of segmenting small breast tumors, Shareef et al. used multi-scale convolutional modules with shared weights to capture the features of breast lesions from ultrasound images (Shareef et al., 2020). To adaptively capture the feature information of breast tumors under different receptive fields, Byra et al. developed a selective kernel U-net (SKU-net) to segment breast tumors using the SK block (Byra et al., 2020). Luo et al. used a channel attention module to calibrate the features extracted from two parallel networks for automatic diagnosis of breast tumors (Luo et al., 2022). Lyu et al. designed an improved pyramid attention network combining attention mechanism and multi-scale features (AMS-PAN) for breast ultrasound image segmentation (Lyu et al., 2023). Inspired by Att U-net, residual learning and multi-scale convolution strategies are added to Att U-net to improve the segmentation precision of lesion tissue. Specifically, a residual convolution block was designed to replace the original convolution module by Tong et al. (Tong et al., 2021). Zhuang et al.

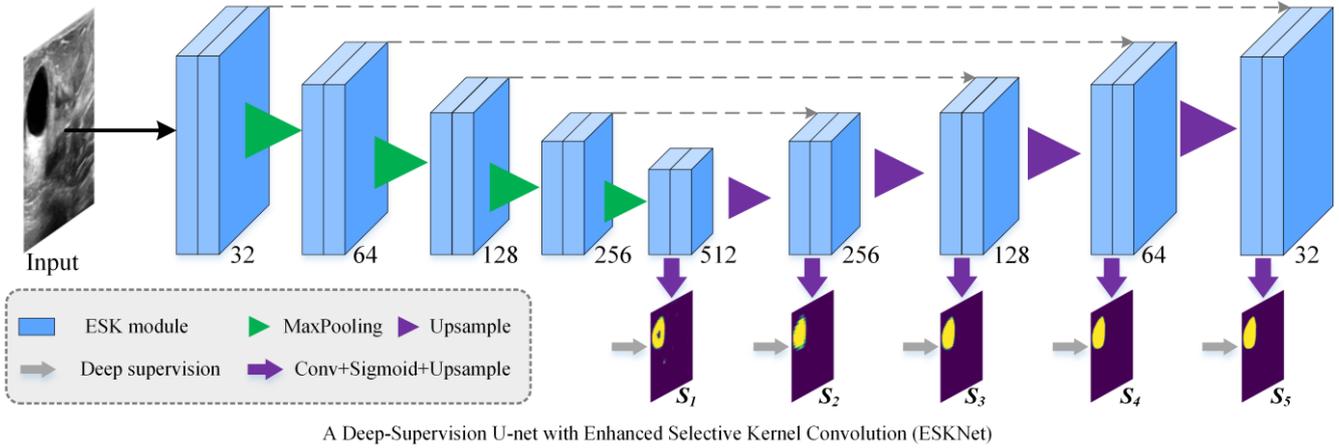

Fig. 3. The illustration of the deep supervision U-net with enhanced selective kernel convolution (ESKNet).

designed a residual block by dilated convolution with different dilation rates to replace the original convolution module (Zhuang et al., 2019). Moon et al. developed a CAD system integrating multiple CNN architectures to automate the diagnosis of breast ultrasound images (Moon et al., 2020). However, this method is limited by the segmentation performance of existing networks on breast ultrasound images. Wang et al. used deep supervision strategy constraints on the feature maps captured at each stage of U-net to segment breast lesions (Wang et al., 2019). With the introduction of the deep supervision component further improves the performance of the network. In this work, we also explored a novel U-shaped framework with deep supervision to segment breast lesions, in which the deep supervision mechanism was only added to the decoding stage.

**2.2 Attention mechanism**

Inspired by human visual attention, many attention algorithms have been developed to boost the CNN's ability to capture the characterization of the objective (Guo et al., 2022). Oktay et al. developed a spatial attention component to weigh low-level semantic feature maps and high-level instance feature maps, which has been successfully applied (Oktay et al., 2018). An architectural component called squeeze ＆ exception (SE) block, which can be calibrated the input feature maps to select out useful channels that better characterize the objective, was developed by Hu et al. (Hu et al., 2020). However, these architectures often tend to learn feature information in fixed convolutional patterns during the encoding and decoding stages (Roy et al., 2018). According to the SE block, Roy et al. designed a novel scSE block that simultaneously squeeze the feature maps along the spatial and the channel dimensions (Roy et al., 2018). Similarly, Zhong et al. developed a new squeeze-and-attention network (SANet) for segmentation tasks based on SE block (Zhong et al., 2020). Hatamizadeh et al. proposed a new transformers-based medical image segmentation model (UNETR), which directly connects the coded representation of transformers to the decoder by using skip-connections (Hatamizadeh et al., 2022). Cao et al. constructed a encoder-decoder architecture (Swin-Unet) with skip-connections based on swin-transformer block for medical image segmentation (Cao et al., 2022). Xu et al. proposed a multi-scale self-attention network (MSSA-Net) to achieve better segmentation accuracy on small datasets (Xu et al., 2021). Zhao et al. developed a novel segmentation network equipped with a focal self-attention block for improving the performance of breast lesion segmentation (Zhao et al., 2022). While many improvement works have considered the improvement of network segmentation performance by the spatial and the channel feature fusion, these methods cannot adaptively select useful features under different receptive fields. In order to alleviate this challenge, Li et al. designed a selective kernel convolution (SK) to adaptively select useful feature information at different scales from the channel dimension (Li et al., 2019). In our work, we introduce an enhanced selected kernel convolution (ESK), which can simultaneously calibrate the features under different receptive fields from the dimensions of the spatial and the channel to upgrade the represented ability.

## 3. Method

Fig. 3 illustrate the deep supervision U-net with enhanced selective kernel convolution (ESKNet) to be implemented in the breast lesion segmentation. Similar to the core structure of U-net (Ronneberger et al., 2015), four down sampling, four up sampling and four skip connection operations are used in our ESKNet. The largest separates are that an enhanced selective kernel convolution module (ESK) is introduced to instead of the original convolution layer in U-net to obtain more precise prediction masks of breast lesions from ultrasound images. ESK contains convolutional layers with different convolutional kernel sizes, which can supply more scales of receptive fields. Capturing objective features from different scales of receptive fields is powerful to strengthen the generalization ability and robustness of the network. As shown in Fig. 2(b), channel calibration and spatial calibration performed concurrently in ESK can help the network capture a more robust characterization of the lesion region from complex breast ultrasound images. In addition, the residual connection in each ESK module can enhance the relevance of long-distance feature information and further improve the segmentation efficiency of the network. To further refine the segmentation results, we use ground-truth masks to constrain each decoding stage.

## 3.1 Enhanced selective kernel convolution (ESK)

The original selection kernel convolution module can be roughly regarded as a multi-scale convolution block with different kernel sizes and a channel attention block, as shown in Fig. 2(a). The developed enhanced selection kernel convolution module is composed of three key components: a multi-scale convolutional block with different convolutional kernel dimensions, a channel-attention module, and a spatial-attention module. In greater detail, the given intermediate feature maps $F \in \mathbb{R}^{c \times h \times w}$ first undergo two parallel convolution operations. The two parallel convolutional operations are a convolutional operation with kernel $5 \times 5$ and a dilated convolutional operation with kernel $3 \times 3$, respectively. The dilation rate of the dilation convolution operation is set as 3. The feature map re-extracted from the two convolution operations is represented as:

$$F_1 = W_{5 \times 5} \times F \quad (1)$$
$$F_2 = W_{3 \times 3} \times F \quad (2)$$

where $F \in \mathbb{R}^{c \times h \times w}$ denotes the given intermediate feature maps, $W_{5 \times 5}$ and $W_{3 \times 3}$ denotes the matrix of the convolutional operation with kernel $5 \times 5$ and the dilated convolutional operation with kernel $3 \times 3$, respectively. $F_1 \in \mathbb{R}^{c \times h \times w}$ and $F_2 \in \mathbb{R}^{c \times h \times w}$ indicates the feature map extracted through the convolutional calculation with kernel $5 \times 5$ and the dilated convolutional calculation with kernel $3 \times 3$, respectively. Subsequently, $F_1 \in \mathbb{R}^{c \times h \times w}$ and $F_2 \in \mathbb{R}^{c \times h \times w}$ are integrated along the dimensions of the channel and input to the channel attention module (See section 3.2) and the spatial attention module (See section 3.3). The merged feature maps can be represented as:

$$F_M = F_1 \oplus F_2 \quad (3)$$

where $\oplus$ denotes the element-wise summation. The feature map $F_M \in \mathbb{R}^{c \times h \times w}$ are performed for channel and spatial dimensions calibration, respectively. Finally, the feature map calibrated by channel and spatial attention are fused with the initial feature maps $F \in \mathbb{R}^{c \times h \times w}$ to obtain a new set of feature maps $F_{CS} \in \mathbb{R}^{c \times h \times w}$:

$$F_{CS} = F \oplus (SAM(F_M) \oplus CAM(F_M)) \quad (4)$$

where $\oplus$ denotes the element-wise addition. $SAM(\cdot)$ is defined as the introduced spatial attention module. $CAM(\cdot)$ is expressed as the designed channel attention module.

## 3.2 Channel attention module

According to Fig. 2(a) and Fig. 2(b), we could observe that the channel attention used in this paper is the same as the channel attention in the SK module. The channel attention can help the network select more useful lesion features through the calibration of channel dimensions. Specifically, the channel-wise statistics are first obtained through a global average pooling (GAP) manipulation. The acquired feature map is expressed as:

$$S_C = GAP(F_M) = \frac{1}{H \times W} \sum_{i=1}^{H} \sum_{j=1}^{W} F_M(i,j) \quad (5)$$

Then, the feature map $S_C$ is performed by the full convolution operation, the batch normalization operation and the ReLU activation operation to produce a new set of feature maps:

$$Z_C = \delta_r(B(W_{fc} \cdot S_C)) \quad (6)$$

where $W_{fc} \in \mathbb{R}^{32 \times 1}$ represents the matrix of full convolution operation, the size of the matrix dimension is 32. $B(\cdot)$ is the batch normalization operation. $\delta_r(\cdot)$ represents the ReLU activation operation. We perform a full convolution operation on the feature map $Z_C$ again for achieving the new feature map:

$$Z = W_{fc} \cdot Z_C \tag{7}$$

where $W_{fc} \in \mathbb{R}^{C \times 1}$ represents the matrix of fully connected (FC) layers, the size of the matrix dimension is $C$. Finally, the feature map $Z$ is executed as the sigmoid operation to get the activation map of the channel attention model:

$$\beta = \sigma_s(Z) \tag{8}$$

In this paper, $\beta \in [0,1]^{c \times 1 \times 1}$ is considered as the activation map for performing channel calibration on $F_2$. Similarly, $1 - \beta \in [0,1]^{c \times 1 \times 1}$ is regarded as the activation map for performing channel calibration on $F_1$. Each value in $\beta$ and $1 - \beta$ can indicate the significance of the channel information on the corresponding volume in $F_1 / F_2$. The feature maps obtained after $F_1$ and $F_2$ are calibrated by the channel activation maps can be expressed as:

$$F_{C1} = (1 - \beta) \cdot F_1 \tag{9}$$
$$F_{C2} = \beta \cdot F_2 \tag{10}$$

where $F_{C1} \in \mathbb{R}^{c \times h \times w}$ and $F_{C2} \in \mathbb{R}^{c \times h \times w}$ are the final outputs of the channel attention module. During the calibration of channel features, the channel activation map is performing the reshaping operation.

### 3.3 Spatial attention module

The calibration on the channel clarifies the importance of each channel, but does not emphasize the location of the objective. Woo et al. pointed out that it is more beneficial to extract useful features to perform calibration operations simultaneously on the dimensions of channels and spaces (Woo et al., 2018). In this work, we developed a spatial attention mechanism based on the original SK module to calibrate spatial features, as shown in Fig. 2(b). It is similar to the channel attention component in the SK module, the spatial attention component we designed also includes two branches. Especially, the feature map $F_M$ is executed with the ReLU operation, the convolution operation with kernel size $1 \times 1$ and the sigmoid operation to produce the activation map on spatial dimension:

$$\alpha = \sigma_s(W \cdot \delta_r(F_M)) \tag{11}$$

where $\delta_r(\cdot)$ and $\sigma_s(\cdot)$ indicate the ReLU operation and sigmoid operation, respectively. $W \in \mathbb{R}^{1 \times h \times w}$ represents the matrix of $1 \times 1$ convolution. In this paper, $\alpha \in [0,1]^{1 \times h \times w}$ is considered as the activation map for performing spatial calibration on $F_2$. Similarly, $1 - \alpha \in [0,1]^{1 \times h \times w}$ is regarded as the activation map for performing channel calibration on $F_1$. Each value in $\alpha$ and $1 - \alpha$ can indicate the significance of the channel information on the corresponding volume in $F_1 / F_2$. The feature maps obtained after $F_1$ and $F_2$ are calibrated by the spatial activation maps can be expressed as:

$$F_{S1} = (1 - \alpha) \cdot F_1 \tag{12}$$
$$F_{S2} = \alpha \cdot F_2 \tag{13}$$

where $F_{S1} \in \mathbb{R}^{c \times h \times w}$ and $F_{S2} \in \mathbb{R}^{c \times h \times w}$ are the final outputs of the spatial attention module. During the calibration of spatial features, the spatial activation map is performing the reshaping operation. The feature map, which are also calibrated by the channel and spatial activation maps, is fused with the input feature maps and fed to the next stage of processing. Finally, the merged feature map can be expressed as:

$$F_{CS} = F \oplus F_{C1} \oplus F_{C2} \oplus F_{S1} \oplus F_{S2} \tag{14}$$

where $\oplus$ denotes the element-wise addition. $F_{CS} \in \mathbb{R}^{c \times h \times w}$ denotes the feature map obtained by the enhanced selection kernel convolution module.

### 3.4 Deep supervision

To make the segmentation mask of breast lesions more similar to the ground truth mask, the depth supervision strategy is added to refine the features captured in the decoding phase, as shown in Fig. 3. By introducing deep supervision constraints, the decoder can learn to generate more accurate segmentation results. Specifically, the feature maps captured at each decoding stage are first processed by the convolution operation with kernel size $1\times1$. Then, the sigmoid operation is used to predict the segmentation results. Finally, the predicted segmentation masks are up-sampled for comparison with reference breast lesion masks. According to the description in Fig. 3, our method can predict five segmentation masks, which can be expressed as:

$$S_i = U^{16/i}(\sigma_s(W \cdot F_{Di})) \quad (15)$$

where $S_i$ represents the predicted mask of the $i-th$ decoding stage, $W \in \mathbb{R}^{1\times h\times w}$ represents the matrix of $1\times1$ convolution. $U^{16/i}(\cdot)$ and $\sigma_s(\cdot)$ represent the up-sampled operation and sigmoid activation operation, respectively. Although the four breast lesion masks predicted in the decoding stage are executed as up-sampling operations, the last segmentation result $S_5$ has a higher accuracy and therefore it is defined as the final segmentation result of our network.

### 3.5 Loss function

In this work, the generally employed BCE function is applied as the loss of our segmentation network. The loss of the method can be denoted as:

$$\mathcal{L} = \sum_{i=1}^{5} \ell_{BCE}^i \quad (16)$$

where $\ell_{BCE}^i$ denote the segmentation loss of the *i-th* decoding stage.

## 4. Materials and Experiments

### 4.1 Datasets

In this paper, three widely used public breast ultrasound datasets are used to evaluate the segmentation network performance. Al-Dhabyani et al.'s 780 breast ultrasound images acquired using two ultrasound machines are adopted as the first dataset for this study (indicates BUSI) (Al-Dhabyani et al., 2020). These breast ultrasound images from the LOGIQ E9 ultrasound and LOGIQ E9 Agile ultrasound system of Baheya Hospital. The average image size of these images is $500\times500$ pixels. These images are classified into three categories: normal, benign, and malignant. The general purpose of breast lesion segmentation in the clinical usage is mainly for the lesion assessment, tracking the lesion change, and identifying distribution and seriousness of lesions (Xue et al., 2021). Therefore, the normal cases in that study were not involved in the training and testing of the network. Yap et al.'s 163 breast ultrasound images collected at two hospitals are chosen as the second dataset for this work (denotes Dataset B) (Yap et al., 2020). These breast ultrasound images were collected in 2012 from the UDIAT Diagnostic Centre of the Corporacio Parc Tauli, Sabadell (Spain) with a Siemens ACUSON Sequoia C512 system. The mean image size of these images is $760\times570$ pixels, where each of the images presented one or more lesions. These images are classified into two categories: benign lesion, and cancerous masse. The third public breast ultrasound dataset is the 42 breast ultrasound images available from Zhuang et al. (Zhuang et al., 2019) (represents STU). These breast ultrasound images were collected from the Imaging Department, First Hospital of Medical College of Shantou University with a GE Voluson E10 ultrasonic diagnostic system. The mean image size of these images is $128\times128$ pixels. These images were not differentiated between benign and malignant lesions. A more detailed description of the three public datasets is given in Table 2.

Table 2 Sample distribution of the three public BUS datasets.

| | Equipment | Benign | Malignant | Normal | Total | Cross experiments | External experiments |
|---|---|---|---|---|---|---|---|
| BUSI | LOGIQ E9 and LOGIQ E9 Agile | 437 | 210 | 133 | 780 | True | False |
| Dataset B | Siemens ACUSON Sequoia C512 | 110 | 53 | No | 163 | True | False |
| STU | GE Voluson E10 | Unknown | Unknown | No | 42 | False | True |

## 4.2 Experimental Settings

The k-fold cross-validation is adopted as the basis for the experimental analysis of this study. During the ablation and comparison experiments, we performed four-fold cross-validation training on BUSI and dataset B, respectively. In the robustness analysis of the network, we executed four-fold and three-fold cross-validation training on benign and malignant breast lesions, respectively. Finally, STU is used to conduct external validation experiments on the already trained segmentation networks. Since the STU only contains 42 images. Therefore, we do not implement cross-validation training on STU.

During training and testing, all the images were resized to have the same size of $384 \times 384$. In the training process, we performed data augmentation on the training data for each fold. In this study, we employed data augmentation methods include random rotation ([$0°$, $20°$] and [$340°$, $357°$]), flipping (along the x-axis and y-axis), elastic transform (alpha=10, sigma=2, alpha_affine=2, random_state=None) , adding Gaussian noise (mean = 0, δ = [5, 10]), blurring (the size of blur kernel is 3 × 3) and random gamma transform (gamma=1.0). The training data of each fold was enlarged 20 times by executing the data augmentation operations. The source code for the data augmentation is available on our public resources (Chen et al., 2023b). For comparison experiments, the methods compared in this paper were retrained on the same training data to ensure that they achieved optimal performance.

In the experiment, Adam with default hyper-parameters is defined as the optimizer of the network. For training, the network's default learning rate is determined to 0.001, and every 10 epochs are reduced by 50%. When the value of the learning rate is smaller than le-4, we assign the learning rate of the network to le-4. During training, we randomly select 20% from each fold of the training data as validation data. We judge whether to terminate the training of the network according to the loss of the validation dataset. After several comparisons, 50 and 12 are chosen as the values for epoch and batch, respectively. The open-source TensorFlow-2.6.0 is used to structure our segmentation network. Our development language and accelerator are Python 3.6 and NVIDIA RTX 3090, respectively.

## 4.3 Evaluation metrics

In this work, five widely-used segmentation metrics are used to quantitatively evaluate the performance of different methods of breast ultrasound image segmentation. They are Jaccard, Precision, Recall, Specificity and Dice (Wang et al., 2020). The mathematical expressions of PA, IoU, Precision, Recall, Specificity and F1 are shown from Eqs. (17) - (21).

$$Jaccard = \frac{TP}{FP+TP+FN} \times 100 \quad (17),$$

$$Precision = \frac{TP}{TP+FP} \times 100 \quad (18),$$

$$Recall = \frac{TP}{TP+FN} \times 100 \quad (19),$$

$$Specificity = \frac{TN}{TN+FP} \times 100 \quad (20),$$

$$Dice = 2 * \frac{Precision \cdot Recall}{Precision + Recall} \quad (21),$$

where TP is defined as the positive output for the corresponding positive ground truth (GT). FP is defined as the positive output for the corresponding negative GT. TN means the negative output for the corresponding negative GT. FN means the negative output for the corresponding positive GT. The higher the value of the five indicators of Jaccard, Precision, Recall, Specificity, and Dice, the better the segmentation result of the network.

## 5. Experimental results

### 5.1 Ablation experiments

Table 3 The quantitative evaluation of our network components on Dataset B and BUSI.

| Components | Types | BUSI Jaccard | Precision | Recall | Specificity | Dice | Dataset B Jaccard | Precision | Recall | Specificity | Dice |
|---|---|---|---|---|---|---|---|---|---|---|---|
| Baseline U-net | All | 60.70±2.36 | 71.88±2.41 | 76.30±2.48 | 96.18±0.55 | 70.10±2.20 | 58.44±4.26 | 70.27±6.11 | 75.32±2.85 | 98.44±0.40 | 68.20±4.23 |
| | Benign | 65.62±2.55 | 78.52±2.61 | 76.57±2.82 | 97.96±0.48 | 73.97±2.20 | 63.31±7.66 | 75.76±4.70 | 76.01±3.79 | 99.09±0.35 | 72.68±6.28 |
| | Malignant | 50.45±2.97 | 58.02±3.06 | 75.78±3.28 | 92.46±1.38 | 62.04±2.97 | 57.15±7.34 | 66.83±11.53 | 77.44±5.19 | 97.53±1.03 | 66.21±7.64 |
| U-net with SK module | All | 68.10±1.63 | 78.62±1.66 | 79.53±1.93 | 97.33±0.45 | 76.92±1.57 | 64.25±4.01 | 75.27±6.70 | 79.36±2.50 | 98.68±0.39 | 73.53±4.05 |
| | Benign | 70.43±2.89 | 81.56±2.44 | 79.56±2.40 | 98.35±0.34 | 78.21±2.77 | 72.06±5.33 | 83.91±4.92 | 83.25±4.00 | 99.46±0.21 | 81.45±4.98 |
| | Malignant | 61.28±1.51 | 71.69±3.13 | 78.04±5.10 | 95.20±1.61 | 71.83±1.63 | 59.79±6.04 | 73.38±12.60 | 75.28±1.70 | 98.02±1.19 | 69.38±6.56 |
| U-net with ESK module | All | 69.67±2.10 | 79.26±1.99 | 81.47±2.97 | 97.40±0.42 | 78.08±2.05 | 67.53±2.38 | 77.25±2.89 | 79.93±3.86 | 98.77±0.25 | 76.34±1.97 |
| | Benign | 73.06±3.12 | 82.53±2.94 | 82.63±2.21 | 98.40±0.41 | 80.42±2.79 | 73.63±9.65 | 82.21±7.89 | 83.37±8.01 | 99.29±0.32 | 81.29±8.28 |
| | Malignant | 62.61±2.85 | 72.45±0.48 | 79.09±6.43 | 95.21±0.98 | 73.23±3.25 | 64.99±5.36 | 73.52±5.17 | **82.00±5.22** | 97.77±1.02 | 74.81±5.77 |
| U-net with ESK module + Deep supervision | All | **70.20±2.28** | **79.57±1.65** | **82.41±2.84** | **97.47±0.35** | **78.71±2.37** | **71.65±2.39** | **81.01±3.91** | **82.66±1.40** | **99.01±0.35** | **79.92±2.21** |
| | Benign | **73.69±2.86** | **82.65±2.75** | **83.89±2.68** | **98.40±0.36** | **81.26±2.88** | **74.61±4.45** | **84.43±7.77** | **83.55±1.82** | **99.47±0.26** | **82.09±4.47** |
| | Malignant | **62.94±2.41** | **73.13±1.02** | **79.35±4.84** | **95.30±0.98** | **73.40±2.52** | **68.98±7.59** | **78.87±8.71** | 81.90±3.43 | **98.15±1.16** | **78.44±7.23** |

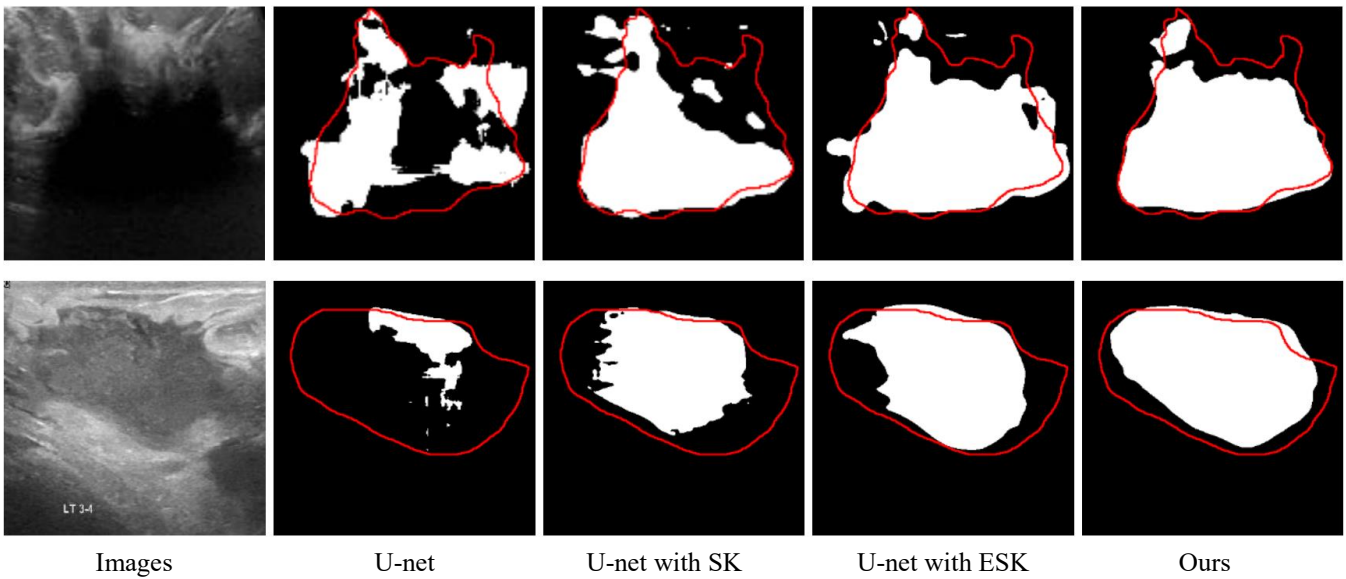

Images  U-net  U-net with SK  U-net with ESK  Ours

Fig. 4. The segmentation results of different network components. With the introduction of SK, ESK, Deep supervision components can gradually improve the segmentation accuracy of breast ultrasound images. Compared to the SK module, the network performance improvement with the introduction of ESK and Deep supervision components is essential.

For fully verify the segmentation effectiveness of network components (such as the SK module, ESK module, and deep supervision), we conduct the four-fold cross-validation training on BUSI and Dataset B, respectively. In the experiment, the reference network is the original U-net. Table 3 illustrates in detail the quantitative evaluation indexes of our components on both Dataset B and BUSI. The visual segmentation results after adding different network components to the U-net are shown in Fig. 4. Based on the results presented in Table 3 and Fig. 4, several conclusions could be drawn. First, the introduction of these components greatly boosted the segmentation capability for breast lesions by the original U-net on breast lesions. Second, according to the comparison between the SK module and ESK module, it can be concluded that the calibration of spatial dimension features through spatial attention is conducive to more robust feature selection. Third, the addition of the deep supervision mechanism enables to urge the segmentation network to output more precise prediction masks of the lesion tissue. To sum up, these key network components developed in this study perform an essential function in improving the segmentation precision of lesion tissue.

Table 4 The comparison of quantitative values (mean ± std) on BUSI and Dataset B. The top 2 scores are marked with bold and blue text, respectively. Asterisks means that the P-value of t-test is less than 0.05.

| Methods | Types | BUSI | | | | | Dataset B | | | | |
|---|---|---|---|---|---|---|---|---|---|---|---|
| | | Jaccard | Precision | Recall | Specificity | Dice | Jaccard | Precision | Recall | Specificity | Dice |
| U-net | All | 60.70±2.36 | 71.88±2.41 | 76.30±2.48 | 96.18±0.55 | 70.10±2.20 | 58.44±4.26 | 70.27±6.11 | 75.32±2.85 | 98.44±0.40 | 68.20±4.23 |
| | Benign | 65.62±2.55 | 78.52±2.61 | 76.57±2.82 | 97.96±0.48 | 73.97±2.20 | 63.31±7.66 | 75.76±4.70 | 76.01±3.79 | 99.09±0.35 | 72.68±6.28 |
| | Malignant | 50.45±2.97 | 58.02±3.06 | 75.78±3.28 | 92.46±1.38 | 62.04±2.97 | 57.15±7.34 | 66.83±11.53 | 77.44±5.19 | 97.53±1.03 | 66.21±7.64 |
| STAN | All | 64.10±3.05 | 73.96±3.30 | 78.39±2.16 | 96.64±0.67 | 73.04±2.95 | 57.09±3.92 | 67.71±3.11 | 69.95±6.17 | 98.58±0.47 | 66.06±4.24 |
| | Benign | 68.18±2.79 | 78.75±3.21 | 79.31±1.18 | 98.04±0.55 | 76.09±2.56 | 60.07±8.68 | 72.80±12.18 | 69.94±7.06 | 99.13±0.45 | 68.93±9.02 |
| | Malignant | 55.62±3.79 | 64.00±3.58 | 76.55±5.40 | 93.71±1.49 | 66.72±3.97 | 62.10±8.39 | 71.78±12.00 | 77.11±4.11 | 97.81±1.12 | 70.93±8.41 |
| Att U-net | All | 57.09±1.22 | 78.78±4.67 | 66.97±4.08 | 96.87±0.83 | 67.99±1.18 | 59.93±4.53 | 70.40±6.05 | 76.15±4.21 | 98.43±0.33 | 69.30±4.07 |
| | Benign | 59.10±1.24 | 82.43±3.85 | 66.00±3.69 | 98.27±0.58 | 69.32±0.50 | 64.78±6.82 | 75.62±5.43 | 78.14±5.70 | 99.11±0.37 | 73.73±5.93 |
| | Malignant | 52.95±3.48 | 69.11±6.47 | 69.03±6.37 | 93.96±1.72 | 65.24±3.89 | 58.21±7.60 | 64.83±8.45 | 77.79±8.35 | 97.35±0.86 | 67.30±8.43 |
| RDAU-net | All | 63.75±3.36 | 71.25±4.11 | 78.90±1.35 | 96.63±0.76 | 71.94±3.46 | 58.17±4.91 | 70.49±4.26 | 73.55±5.28 | 98.37±0.39 | 68.22±4.94 |
| | Benign | 67.64±3.27 | 75.50±3.89 | 79.35±2.47 | 98.00±0.34 | 74.84±3.40 | 59.53±5.99 | 71.51±4.22 | 71.69±7.43 | 98.93±0.37 | 68.35±6.07 |
| | Malignant | 55.65±3.68 | 62.38±4.94 | 78.00±2.80 | 93.78±1.90 | 65.90±3.85 | 60.89±6.82 | 72.55±7.63 | 77.55±4.37 | 97.57±2.98 | 71.62±7.27 |
| MADU-net | All | 61.62±2.69 | 73.77±2.90 | 76.87±2.58 | 96.40±0.62 | 71.35±2.67 | 63.09±3.04 | 73.70±5.08 | 79.24±1.72 | 98.61±0.36 | 72.32±3.14 |
| | Benign | 65.42±2.49 | 79.22±2.61 | 76.32±2.03 | 97.96±0.52 | 74.21±2.41 | 68.82±9.05 | 80.89±9.28 | 80.92±5.18 | 99.28±0.35 | 77.74±8.81 |
| | Malignant | 53.71±3.64 | 62.44±3.71 | 78.04±4.04 | 93.14±1.33 | 65.40±3.70 | 61.06±7.59 | 67.27±9.08 | 81.25±4.58 | 97.54±1.16 | 69.76±7.44 |
| U-net++ | All | 61.38±1.73 | **79.68±3.07** | 71.44±2.77 | 97.04±0.54 | 71.58±2.09 | 61.19±5.86 | 68.32±5.73 | 79.64±3.84 | 98.44±0.41 | 69.77±5.30 |
| | Benign | 63.79±1.68 | **84.24±2.95** | 70.39±2.11 | 98.37±0.48 | 73.26±1.79 | 67.84±7.56 | 75.27±7.34 | 83.18±5.56 | 99.10±0.40 | 76.43±7.60 |
| | Malignant | 56.39±3.40 | 70.18±3.31 | 73.68±5.75 | 94.28±1.08 | 68.09±3.95 | 58.73±10.76 | 65.30±12.31 | 79.52±5.26 | 97.51±1.09 | 67.14±10.82 |
| SegNet | All | 67.31±1.87 | 76.09±2.00 | 80.85±1.03* | 96.99±0.53 | 75.64±1.80 | 62.83±2.20 | 71.72±1.70 | 80.15±3.90* | 98.59±0.30 | 72.16±1.52 |
| | Benign | 71.18±2.44* | 80.70±2.51 | 80.99±2.23 | 98.24±0.40 | 78.43±2.37 | 64.86±3.23 | 74.29±5.33 | 79.20±4.13 | 99.15±0.25 | 73.28±3.34 |
| | Malignant | 59.28±1.40 | 66.48±1.51 | **80.60±2.90** | 94.36±1.31 | 69.86±1.66 | 62.87±7.58 | 70.36±6.47 | **83.32±6.72** | 97.52±0.99 | 73.16±7.79 |
| AE U-net | All | 64.57±2.91 | 74.44±3.74 | 79.00±2.11 | 96.80±0.54 | 73.47±3.03 | 62.37±2.16 | 72.27±1.91 | 78.97±2.29 | 98.67±0.28 | 72.23±2.14 |
| | Benign | 67.91±3.16 | 78.55±4.36 | 79.56±1.17 | 98.16±0.53 | 75.81±3.13 | 63.66±4.91 | 74.61±6.52 | 78.02±3.11 | 99.17±0.31 | 73.39±5.12 |
| | Malignant | 57.62±3.25 | 65.87±2.51 | 77.86±4.89 | 93.95±0.98 | 68.60±3.49 | 64.42±6.42* | 73.13±8.94 | 81.76±1.59 | 97.86±1.15 | 73.86±6.80* |
| UNETR | All | 63.86±1.78 | 73.44±2.85 | 80.21±1.60 | 96.65±0.65 | 72.99±1.67 | 59.11±5.76 | 70.39±3.15 | 78.17±4.31 | 98.04±0.53 | 69.64±5.28 |
| | Benign | 67.38±1.10 | 77.63±2.72 | 80.56±0.50 | 97.94±0.35 | 75.59±1.56 | 61.60±6.72 | 76.26±7.42 | 76.10±6.77 | 98.94±0.62 | 71.85±6.64 |
| | Malignant | 56.56±3.17 | 64.72±3.41 | 79.52±4.08 | 93.96±1.67 | 67.60±2.78 | 60.03±6.21 | 66.62±5.76 | 81.57±6.62 | 96.90±0.86 | 70.52±7.01 |
| AMS-PAN | All | 65.06±1.47 | 77.23±2.34 | 77.69±1.85 | 96.96±0.57 | 74.78±1.50 | 62.36±4.53 | 72.27±5.32 | 79.95±4.50 | 98.59±0.26 | 71.78±3.65 |
| | Benign | 68.30±1.34 | 80.87±1.94 | 78.25±0.99 | 98.19±0.42 | 76.95±1.02 | 67.33±6.53 | 77.33±6.40 | 83.48±5.81* | 99.12±0.31 | 76.77±6.05 |
| | Malignant | 58.34±2.85 | 69.67±3.41 | 76.54±4.72 | 94.39±1.13 | 70.27±3.02 | 59.58±7.56 | 66.33±8.20 | 79.92±7.33 | 97.63±1.12 | 68.32±7.90 |
| SANet | All | 65.96±2.78 | 74.84±4.18 | 80.76±2.30 | 96.75±0.70 | 74.46±2.76 | 63.26±7.77 | 72.40±10.19 | 77.81±5.07 | 98.61±0.57 | 71.84±7.96 |
| | Benign | 69.61±2.15 | 79.22±3.69 | 81.03±1.37* | 98.10±0.56 | 77.10±2.06 | 66.54±4.96 | 76.65±6.18 | 77.62±7.00 | 99.26±0.17 | 74.79±5.78 |
| | Malignant | 58.38±4.43 | 65.75±5.36 | 80.23±4.61 | 93.94±1.24 | 69.00±4.74 | 57.76±8.79 | 66.18±13.19 | 75.84±5.93 | 97.53±1.18 | 66.38±8.96 |
| scSEU-net | All | 67.68±2.28 | 78.95±2.73 | 79.58±1.14 | 97.26±0.48 | 76.67±2.20 | 62.17±5.03 | 71.31±5.44 | 79.16±5.79 | 98.52±0.22 | 71.30±4.24 |
| | Benign | 70.62±2.48 | 81.86±3.33 | 80.08±1.71 | 98.30±0.50 | 78.54±2.37* | 68.03±9.56 | 77.68±8.15 | 82.09±7.24 | 99.14±0.46 | 76.86±8.34 |
| | Malignant | 61.57±2.59* | 72.87±2.04* | 78.61±4.70 | 95.08±1.08 | 72.81±2.48* | 60.21±7.92 | 66.63±10.34 | 80.39±5.71 | 97.62±1.11 | 69.09±8.53 |
| SKU-net | All | 68.10±1.63* | 78.62±1.66 | 79.53±1.93 | 97.33±0.45 | 76.92±1.57* | 64.25±4.01* | 75.27±6.70 | 79.36±2.50 | 98.68±0.39 | 73.53±4.05* |
| | Benign | 70.43±2.89 | 81.56±2.44 | 79.56±2.40 | 98.35±0.34* | 78.21±2.77 | 72.06±5.33* | 83.91±4.92* | 83.25±4.00 | 99.46±0.21* | 81.45±4.98* |
| | Malignant | 61.28±1.51 | 71.69±3.13 | 78.04±5.10 | 95.20±1.61* | 71.83±1.63 | 59.79±6.04 | 73.38±12.60* | 75.28±1.70 | 98.02±1.19* | 69.38±6.56 |
| Ours | All | **70.20±2.28** | 79.57±1.65 | **82.41±2.84** | **97.47±0.35** | **78.71±2.37** | **71.65±2.39** | **81.01±3.91** | **82.66±1.40** | **99.01±0.35** | **79.92±2.21** |
| | Benign | **73.69±2.86** | 82.65±2.75 | **83.89±2.68** | **98.40±0.36** | **81.26±2.88** | **74.61±4.45** | **84.43±7.77** | **83.55±1.82** | **99.47±0.26** | **82.09±4.47** |
| | Malignant | **62.94±2.41** | **73.13±1.02** | 79.35±4.84 | 95.30±0.98 | **73.40±2.52** | **68.98±7.59** | **78.87±8.71** | 81.90±3.43 | 98.15±1.16 | **78.44±7.23** |

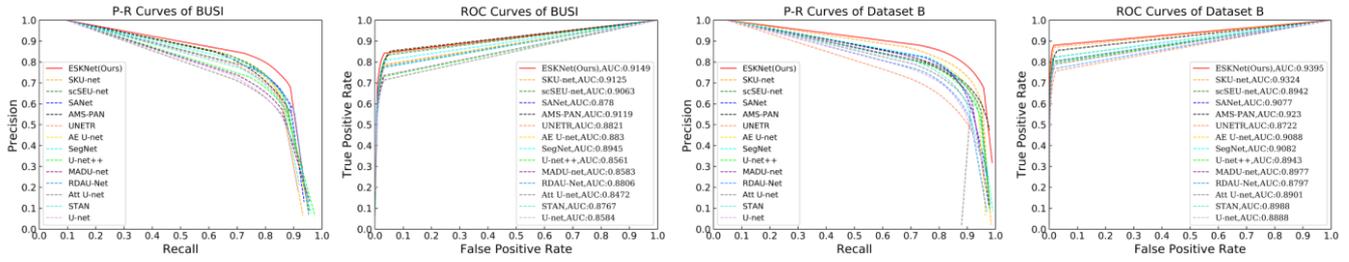

Fig. 5.　The P-R and ROC curves after segmentation for all images of BUSI and all images of Dataset B using various methods.

**5.2　Comparison experiments**

　　In this study, thirteen methods for ultrasound image or medical image segmentation are used to compare with our methods. These classical segmentation methods include U-net (Ronneberger et al., 2015), SegNet (Badrinarayanan et al., 2017), Att U-net (Oktay et al., 2018), U-net++ (Zhou et al., 2020), UNETR (Hatamizadeh et al., 2022), scSEU-net (Roy et al., 2018), SANet (Zhong et al., 2020), RDAU-Net (Zhuang et al., 2019), MADU-net (Abraham and Khan, 2019), STAN (Shareef et al., 2020), SKU-net (Byra et al., 2020), AE U-Net (Yan et al., 2022), AMS-PAN (Lyu et al., 2023). U-Net++ and UNETR are two variants of the U-net network for medical image segmentation. SegNet is also one of successful segmentation methods by recording the location information of the objective. To ensure the fairness of the experiments, all comparison methods are adequately retrained on the same training data. In the comparison experiments, the segmentation results of all methods are not subjected to any post-processing.

　　The evaluation results of BUSI and Dataset B through various segmentation methods are presented in Table 3. The comparison with these segmentation methods indicates that our method holds the ideal segmentation results of breast ultrasound. The values of five indicators on all images of BUSI are 70.20%, 79.57%, 82.41%, 97.47%, and 78.71%, respectively. The values of five indicators on all images of Dataset B are 71.65%, 81.01%, 82.66%, 99.01%, and 79.92%, respectively. Although U-net++ has a higher Precision value on benign tumors in the BUSI dataset, the value on malignant tumors is unsatisfactory. SegNet achieves the best Recall value on malignant tumor segmentation, which suggests that recording down-sampled positional information can mitigate the occurrence of missed detections to some extent. However, based on other metrics, the method also inevitably causes more false detections. In summary, the overall performance of U-net++ and SegNet on BUSI and Dataset B needs to be further improved. For more visual comparison of the different segmentation methods, t-test experiments were performed on the segmentation results. The p-value ($p < 0.05$) of t-test represents our method has significant improvement in five indicators.

　　Fig. 5 shows the P-R and ROC curves after segmentation of all images in BUSI and of all images in Dataset B by various methods. Based on the region of the curve, we can also compare the confidence levels of the different methods on this segmentation task. In addition, we give the values of the different methods on the ACU metric in the ROC curve. The AUC values of our method on BUSI and Dataset B are 0.9149 and 0.9395, respectively. In terms of the curve area and the AUC, our approach achieves the highest results. This demonstrated that our approach achieves the highest confidence level in the segmentation of BUSI and Dataset B.

　　The comparison with these segmentation approaches in five quantitative evaluation indicators fully demonstrates the competitiveness of our method. In addition, the analysis based on P-R curves and ROC curves also shows the feasibility of our method in breast lesion segmentation. In order to compare each segmentation method more intuitively, the prediction masks of breast tumors are visualized. The visual segmentation masks of various methods are shown in Fig. 6. In general, according to Table 4 and Fig. 6, we can draw three key points: (i) Unlike the comparison approach, our approach could better deal with different input images. As shown in Fig. 6, our method is less disturbed by the disease category, lesion morphology, and surrounding tissues. This adequately demonstrates that our approach has good adaptability and can extract robust breast tumor representation. (ii) Although various optimized U-nets improve the segmentation precision of lesion tissue to varying degrees, especially SKU-net, scSEU-net, etc., these variants still have room for further improvement. As shown in Table 4, different segmentation methods have achieved relatively competitive results in individual evaluation indicators, but their overall performance in BUSI and Dataset B is not ideal.

(iii) As you can see in the second and fifth rows, due to the complexity of ultrasound modalities, these variant networks still have seriously missed detections and false detections on individual images, and even the segmentation fails. (iv) Some errors still occur in the segmentation of breast lesions with our method. It is worth noting that the prediction masks of this method has fewer missed detection regions and are closer to the ground-truth mask. The overall performance of this approach on breast tumor segmentation is superior to other algorithms.

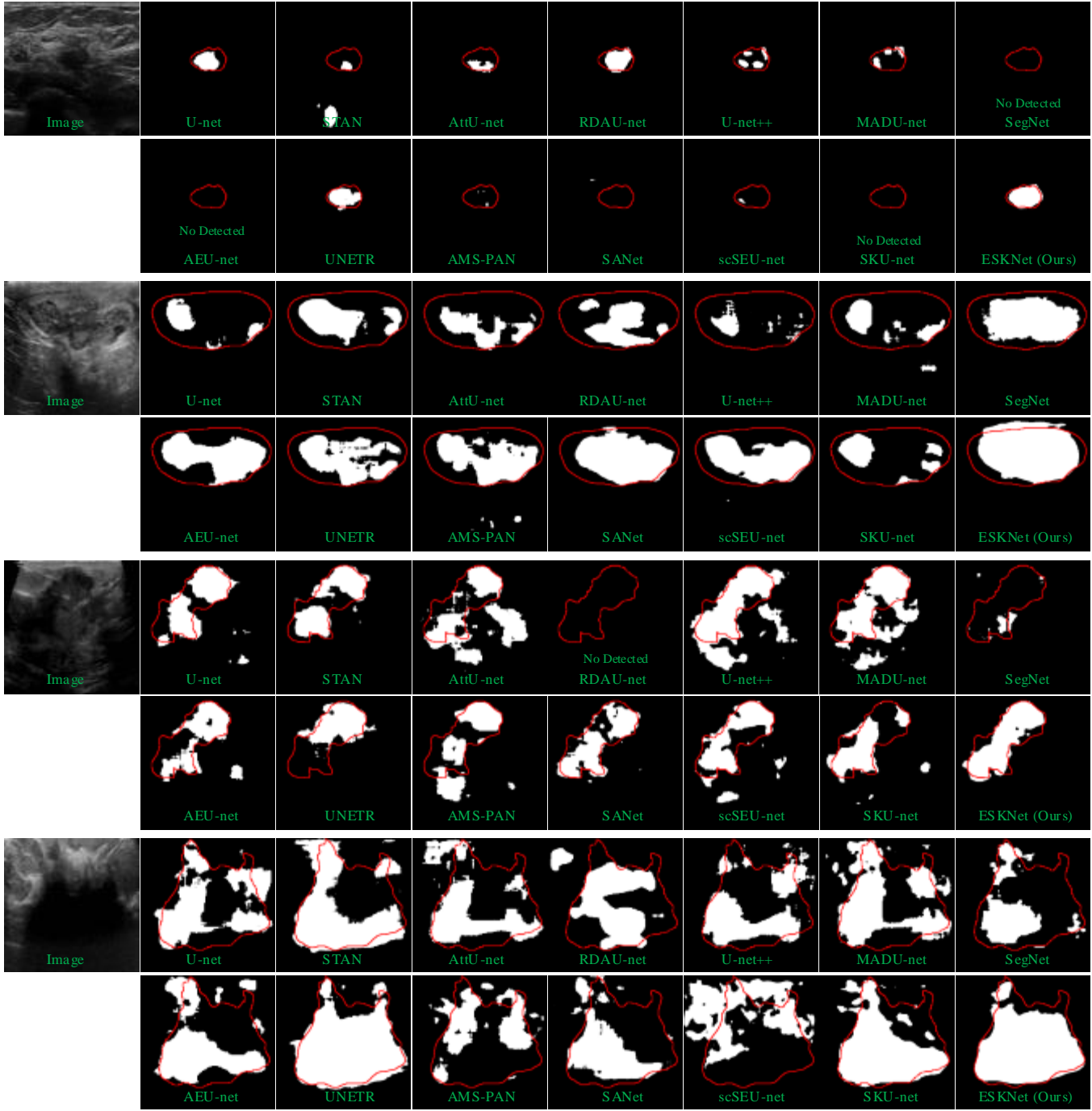

Fig. 6. The prediction masks of BUSI and Dataset B. Some methods fail to segment breast tumors on individual ultrasound images.

## 6. Discussions

The deep supervision U-net with enhanced selective kernel convolution (ESKNet) is developed for breast tumor segmentation. The enhanced selective kernel convolution (ESK) not only adaptively selects features under different scale receptive fields from the channel and spatial dimensions but also further strengthens the relevance of remote feature

14 / 21information. The design of the ESK module can effectively mitigate the interference of different scales and morphologies of breast tumors on the segmentation network and improve the generalization ability. In addition, performing deeply supervised operations on the decoding phase of the network can further motivate the network to predict more complete masks of breast lesions. In this study, we conducted an extensive segmentation performance analysis on three public breast ultrasound datasets. The superiority of this method is well evidenced by comparison with the thirteen segmentation methods. In this section, we further discuss the robustness and limitations of this study.

**Table 5** The comparison of quantitative values (mean ± std) on benign and malignant tumors. The top 2 scores are marked with bold and blue text, respectively. Asterisks means that the P-value of t-test is less than 0.05.

| Methods | Benign tumors of BUSI | | | | | Malignant tumors of BUSI | | | | |
|---|---|---|---|---|---|---|---|---|---|---|
| | Jaccard | Precision | Recall | Specificity | Dice | Jaccard | Precision | Recall | Specificity | Dice |
| U-net | 61.53±3.98 | 74.97±2.80 | 73.97±5.81 | 97.72±0.59 | 70.49±3.23 | 51.11±2.62 | 64.96±2.55 | 68.86±4.27 | 93.63±1.28 | 63.47±2.38 |
| STAN | 64.20±2.73 | 73.37±3.70 | 77.91±1.27 | 97.65±0.52 | 71.98±2.98 | 51.01±2.38 | 62.74±3.72 | 70.96±6.75 | 93.69±1.56 | 62.50±1.97 |
| Att U-net | 65.03±2.05 | 75.24±1.68 | 79.44±2.84 | 97.68±0.62 | 73.30±2.00 | 51.12±2.35 | 61.62±0.97 | 72.57±2.17 | 93.12±1.00 | 62.95±2.14 |
| RDAU-net | 64.70±2.17 | 72.54±1.57 | 79.36±0.98 | 97.79±0.28 | 72.70±1.62 | 51.63±1.62 | 60.85±5.01 | 71.89±2.55 | 93.47±1.45 | 62.44±2.21 |
| U-net++ | 68.25±2.75 | 75.93±3.66 | 81.58±1.09* | 97.74±0.62 | 75.56±2.79 | 54.03±3.03 | 65.50±2.94 | 73.43±2.10 | 93.73±1.31 | 65.52±2.75 |
| MADU-net | 66.74±2.10 | 76.74±2.94 | 79.97±1.64 | 97.75±0.60 | 74.82±2.26 | 54.12±2.96 | 67.46±3.40 | 72.36±5.05 | 93.94±1.25 | 65.77±2.58 |
| SegNet | 67.89±3.31 | 76.96±3.11 | 79.57±2.21 | 97.98±0.46 | 75.47±2.91 | 54.89±1.78 | 63.79±2.65 | **77.25±4.02** | 94.00±1.14 | 65.90±1.97 |
| AE U-net | 67.89±1.96 | 77.17±3.63 | 80.54±1.25 | 97.95±0.60 | 75.77±1.82 | 55.38±1.77 | 67.87±3.81 | 72.88±3.47 | 94.43±1.33 | 66.50±1.52 |
| UNETR | 67.30±1.11 | 77.10±2.08 | 80.57±1.45 | 97.88±0.41 | 75.63±0.94 | 54.17±1.60 | 65.40±5.89 | 75.02±5.18 | 93.94±1.90 | 65.67±1.30 |
| AMA-PAN | 69.21±2.08 | 79.24±2.84 | 81.04±1.00 | 98.02±0.51 | 77.36±1.19 | 55.17±3.51 | 66.78±6.01 | 73.54±4.58 | 94.11±1.66 | 66.75±2.96 |
| SANet | 65.96±2.78 | 74.84±4.18 | 80.76±2.30 | 96.75±0.70 | 74.46±2.76 | 57.55±1.28* | 67.49±0.80 | 75.91±3.91 | 94.21±1.12 | 68.45±1.05* |
| scSEU-net | 71.33±2.45* | 80.76±2.69* | 81.58±2.45 | 98.20±1.47* | 78.97±2.44* | 56.21±2.16 | 67.75±4.77 | 73.99±3.22 | 94.29±1.13 | 67.39±1.96 |
| SKU-net | 69.91±2.11 | 79.15±2.05 | 81.54±2.17 | 98.06±0.52 | 77.88±2.98 | 57.06±2.42 | 69.59±4.20* | 73.58±6.75 | 94.65±1.49* | 68.19±2.28 |
| Ours | **72.73±2.12** | **81.50±2.62** | **82.69±0.40** | **98.29±0.40** | **80.17±1.79** | **59.63±1.57** | **71.52±3.40** | 74.71±3.21 | **95.24±1.31** | **70.43±1.32** |

## 6.1 Robustness Analysis

### 6.1.1 Robustness on Benign and Malignant Lesions

There are great differences between benign tumors and malignant tumors in morphology and intensity distribution. Generally speaking, benign lesions have regular morphology, but the tumor size varies greatly. The shape of malignant lesions is irregular, the internal energy distribution is uneven and the boundary is more blurred. To this end, we conducted comparative experiments on benign and malignant breast tumors, respectively. The four-fold cross-validation experiment is conducted for the benign ultrasound image. The three-fold cross-validation experiment is conducted for the malignant ultrasound image. In Table 5, the quantitative results of benign and malignant lesions obtained using various segmentation methods are displayed. Based on the results presented in Table 5, the sensitivity of the same segmentation method to benign and malignant lesions is different. In the segmentation of benign lesions, the most optimal values on the five evaluation parameters were obtained by our approach. In the segmentation of malignant ultrasound images, our methodology gained the most optimal values on the four parameters of Jaccard, Precision, Specificity, and Dice, and the second value on recall. Compared with the existing breast segmentation techniques, the method can better cope with the segmentation tasks of different lesion types and obtain the most competitive results. The p-values of t-test experiments further demonstrates the robustness and superiority of this approach.

### 6.1.2 External validation

In addition to the influence of the type of breast lesions, there are also great differences between ultrasound images collected by different devices. These differences will also seriously affect the network segmentation performance (tend to fail to achieve ideal segmentation results on external test data). In this paper, STU provided by Zhuang et al. (Zhuang et al., 2019) is adopted as outside test dataset to further judge the capability of each algorithm to segment breast lesions. Specifically, we first performed four-fold cross-validation training on BUSI and dataset B, respectively. Then, we conducted external validation experiments using the STU dataset on each fold of the already trained segmentation network.

Table 6 The comparison of quantitative values (mean ± std) on the external test data STU. The top 2 scores are marked with bold and blue text, respectively. Asterisks means that the P-value of t-test is less than 0.05.

| Methods | STU on BUSI | | | | | STU on Dataset B | | | | |
|---|---|---|---|---|---|---|---|---|---|---|
| | Jaccard | Precision | Recall | Specificity | Dice | Jaccard | Precision | Recall | Specificity | Dice |
| U-net | 62.40±5.06 | 73.11±6.40 | 82.47±2.34 | 95.53±0.85 | 74.09±4.62 | 58.90±3.75 | 66.27±5.46 | 86.88±1.60 | 94.54±0.74 | 71.41±3.67 |
| STAN | 70.92±2.21 | 82.47±1.65 | 84.20±1.86 | 97.09±0.32 | 81.33±1.90 | 58.95±3.31 | 64.38±3.73 | 90.72±0.41 | 94.59±0.57 | 70.76±3.16 |
| Att U-net | 56.44±2.97 | 65.20±4.12 | 84.21±1.09 | 94.35±0.54 | 69.81±2.72 | 52.65±2.29 | 59.26±2.86 | 86.35±1.29 | 93.41±0.41 | 65.19±2.73 |
| RDAU-net | 69.36±4.57 | 79.82±6.29 | 86.07±1.61 | 96.78±1.02 | 79.89±4.01 | 60.11±3.02 | 63.49±2.86 | 91.37±1.07 | 94.70±0.45 | 72.40±2.74 |
| U-net++ | 61.69±3.60 | 70.70±5.02 | 86.01±1.01 | 95.27±0.58 | 73.73±3.59 | 59.18±4.21 | 64.86±5.36 | 89.67±1.59 | 94.33±0.83 | 70.70±4.19 |
| MADU-net | 59.36±1.92 | 69.24±2.44 | 82.36±0.85 | 94.91±0.42 | 72.27±1.65 | 58.11±3.24 | 66.05±3.69 | 86.17±0.99 | 94.35±0.47 | 70.32±3.01 |
| SegNet | 75.55±1.33 | 86.96±0.21 | 86.32±1.40* | 97.93±0.11 | 85.21±0.99 | 62.70±3.09 | 66.57±3.05 | 91.36±0.38 | 95.04±0.49 | 73.50±3.62 |
| AE U-net | 71.92±0.65 | 83.10±0.45 | 84.86±0.84 | 97.35±0.16 | 82.12±0.60 | 62.46±2.44 | 66.79±2.95 | 91.24±0.42 | 95.00±0.39 | 74.41±2.30 |
| UNETR | 69.43±1.52 | 82.89±2.88 | 82.50±1.85 | 97.38±0.43 | 80.78±1.24 | 55.16±1.65 | 59.80±1.95 | 91.87±1.03* | 93.65±0.22 | 67.54±1.45 |
| AMA-PAN | 65.93±1.06 | 81.52±2.66 | 79.80±3.21 | 96.77±0.50 | 78.54±0.81 | 55.35±1.73 | 60.24±1.63 | 91.43±0.54 | 93.57±0.35 | 67.88±2.08 |
| SANet | 69.89±2.12 | 79.58±2.90 | 86.28±1.52 | 96.78±0.30 | 79.62±2.13 | 63.99±5.42 | 69.91±7.58 | 91.03±2.23 | 95.37±0.80 | 74.99±5.16 |
| scSEU-net | 71.97±0.73 | 86.15±0.61 | 83.01±1.18 | 97.71±0.12 | 82.84±0.66 | 55.12±5.83 | 60.87±6.76 | 89.73±1.63 | 93.85±0.60 | 67.30±5.93 |
| SKU-net | 76.23±0.78* | 89.26±1.64* | 84.60±1.27 | 98.31±0.31* | 85.56±0.70* | 66.94±3.15* | 71.99±4.08* | 91.44±1.08 | 95.40±0.51* | 78.29±3.05* |
| Ours | 77.65±0.70 | 90.23±0.93 | 86.49±0.79 | 98.48±0.26 | 86.79±0.56 | 72.63±2.03 | 79.23±3.39 | 92.01±0.81 | 96.57±0.53 | 82.72±1.70 |

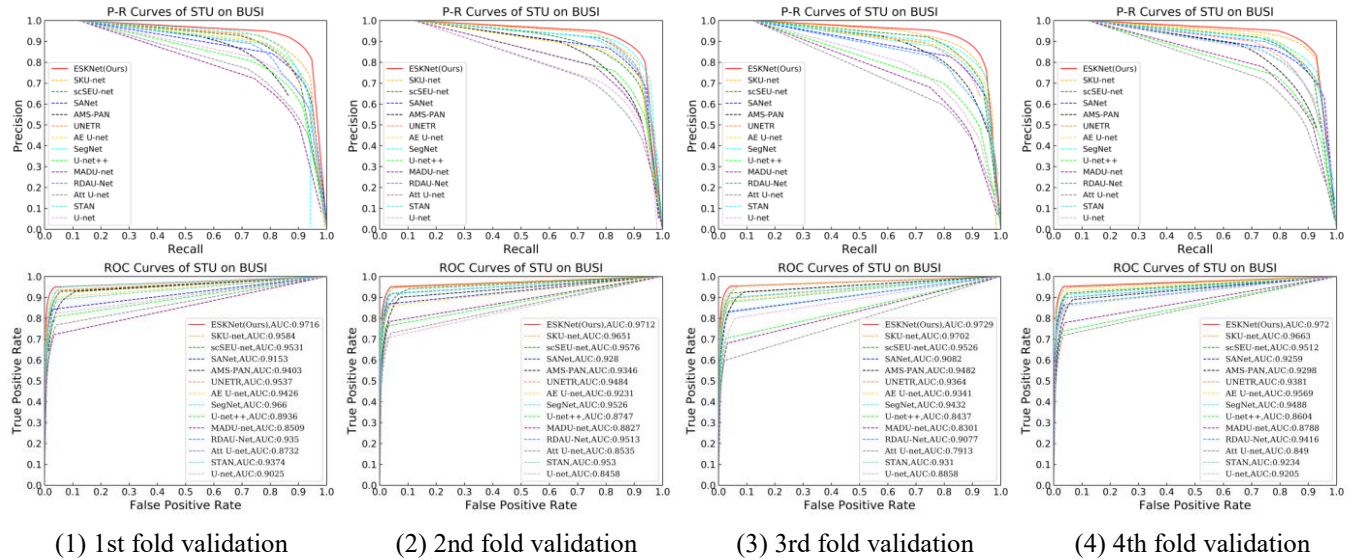

(1) 1st fold validation  (2) 2nd fold validation  (3) 3rd fold validation  (4) 4th fold validation

Fig. 7. P-R and ROC curves for external verification of STU after training on BUSI.

Table 6 shows the evaluation results of external test experiments. For the external test of BUSI and Dataset B, our methodology yielded the most optimal results. In the external experiment of BUSI, the values of the five evaluation indicators are 77.65%, 90.23%, 86.49%, 98.48%, and 86.79%, respectively. The values of the five evaluation indicators on Dataset B are 72.63%, 79.23%, 92.01%, 96.57%, and 82.72%, respectively. Although some methods achieved competitive results in K-fold cross-validation experiments, they did not perform well in external validation data (such as scSEU-net). This adequately demonstrates that the generalization ability and robustness of these methods are limited. Fig. 7 and Fig. 8 shows the P-R and ROC curves for external verification of STU after training on BUSI and Dataset B, respectively. The AUC value for our approach to perform external tests on BUSI is 0.9716, 0.9712, 0.9729 and 0.972, respectively. The AUC value for our approach to perform external tests on Dataset B is 0.9509, 0.9305, 0.9391 and 0.8998, respectively. Based on the area enclosed by the curve and the AUC value, we can draw the conclusion of our method achieves the highest confidence level in the external validation experiment of each fold. According to Table 6, Fig. 7, and Fig. 8, the proposed methodology exhibits superior generalization ability on breast ultrasound image segmentation task.

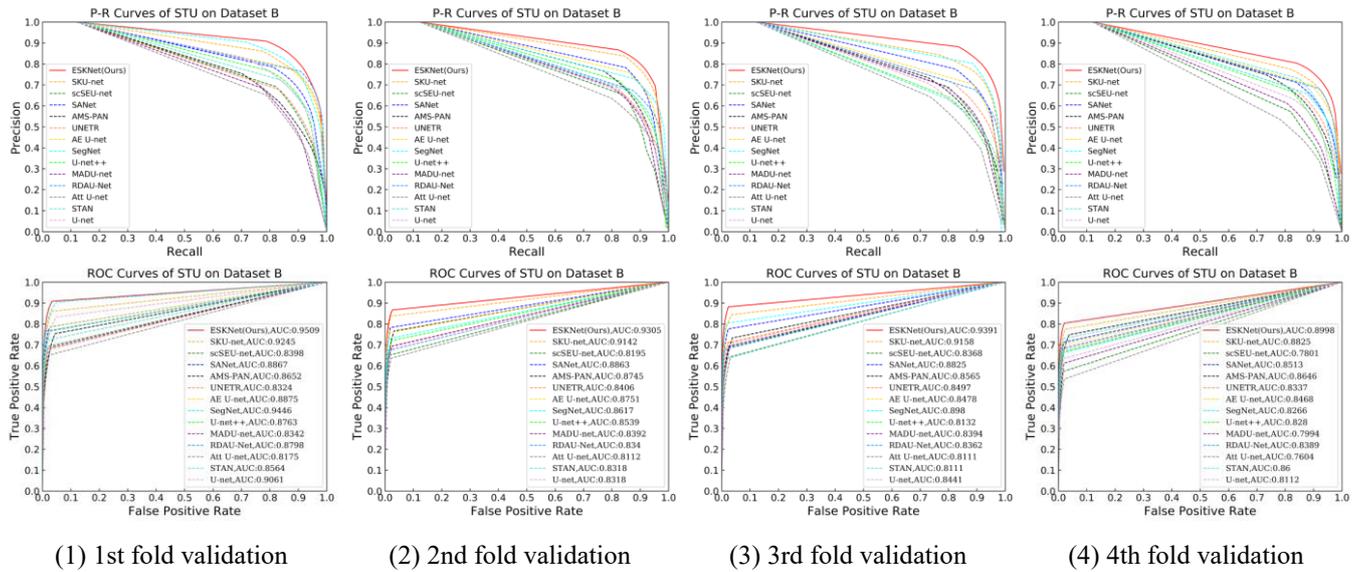

(1) 1st fold validation  (2) 2nd fold validation  (3) 3rd fold validation  (4) 4th fold validation

Fig. 8. P-R and ROC curves for external verification of STU after training on Dataset B.

Table 7 The comparison of quantitative values (mean ± std) on noise degraded ultrasound images. The top 2 scores are marked with bold and blue text, respectively. Asterisks means that the P-value of t-test is less than 0.05.

| Methods | BUSI | | | | | Dataset B | | | | |
|---|---|---|---|---|---|---|---|---|---|---|
| | Jaccard | Precision | Recall | Specificity | Dice | Jaccard | Precision | Recall | Specificity | Dice |
| U-net | 59.19±1.87 | 71.01±1.92 | 75.06±2.38 | 96.19±0.51 | 68.79±1.73 | 58.27±2.92 | 70.00±5.03 | 74.56±3.60 | 98.23±0.42 | 68.13±2.67 |
| STAN | 63.79±3.09 | 73.95±3.54 | 78.17±2.52 | 96.67±0.68 | 72.87±3.13 | 56.58±4.11 | 67.58±3.83 | 69.22±4.21 | 98.41±0.47 | 66.06±4.59 |
| Att U-net | 57.65±1.54 | 69.06±2.13 | 75.05±1.96 | 96.04±0.61 | 67.76±1.58 | 59.08±3.90 | 69.77±4.93 | 76.08±3.76 | 98.36±0.28 | 69.22±3.27 |
| RDAU-net | 63.31±3.39 | 70.83±4.14 | 78.79±1.53 | 96.65±0.77 | 71.60±3.42 | 57.75±4.78 | 70.52±4.48 | 72.37±4.70 | 98.43±0.45 | 67.56±4.89 |
| U-net++ | 61.27±1.35 | 72.07±2.33 | 77.61±2.49 | 96.33±0.58 | 70.85±1.69 | 61.12±4.90 | 68.52±4.69 | 77.59±4.26 | 98.49±0.40 | 69.80±4.14 |
| MADU-net | 59.36±2.20 | 72.26±2.05 | 75.98±2.83 | 96.29±0.52 | 69.60±2.24 | 61.87±2.61 | 73.91±3.89 | 76.17±3.09 | 98.67±0.37 | 71.55±2.31 |
| SegNet | 66.48±1.80 | 75.84±1.57 | 80.06±1.25* | 97.01±0.54 | 75.02±1.72 | 62.20±2.66 | 72.37±3.19 | 78.38±3.37 | 98.61±0.31 | 71.60±3.18 |
| AE U-net | 63.80±2.36 | 73.67±3.13 | 78.45±2.10 | 96.77±0.49 | 72.81±2.52 | 61.74±2.78 | 72.47±1.22 | 78.18±2.85 | 98.70±0.32 | 72.05±2.33 |
| UNETR | 63.18±1.93 | 73.36±2.92 | 78.92±2.06 | 96.38±0.59 | 72.39±1.80 | 58.85±6.54 | 71.93±3.70 | 77.45±5.35 | 98.13±0.58 | 69.43±5.95 |
| AMA-PAN | 62.52±1.59 | 75.82±2.26 | 75.70±1.98 | 96.87±0.57 | 72.72±1.29 | 61.30±4.97 | 69.75±5.63 | 79.40±3.24 | 98.51±0.25 | 70.72±4.05 |
| SANet | 64.99±2.73 | 73.83±3.68 | 80.04±2.17 | 96.75±0.66 | 73.62±2.72 | 62.79±4.57 | 72.84±7.17 | 79.35±3.86 | 98.67±0.50 | 71.95±4.91 |
| scSEU-net | 66.63±2.36 | 78.77±2.64* | 78.44±1.11 | 97.30±0.46 | 75.90±2.21 | 61.60±4.42 | 70.97±5.13 | 78.59±6.03 | 98.53±0.26 | 70.99±3.64 |
| SKU-net | 67.84±1.61* | 78.43±1.74 | 79.92±2.24 | 97.43±0.50 | 76.77±1.54* | 65.90±4.78* | 77.62±6.54* | 79.44±2.96* | 98.90±0.45* | 75.41±4.70* |
| Ours | 69.47±2.10 | 78.85±1.37 | 81.92±2.86 | 97.38±0.40 | 78.09±2.19 | 71.56±2.07 | 81.83±1.59 | 82.24±2.03 | 99.08±0.36 | 80.26±1.58 |

*6.1.3  Robustness on degraded ultrasound images*

Similar surrounding tissues in the ultrasound image can cause blurring of the lesion boundary. In addition, ultrasound images with severe speckle noise. These intrinsic and extrinsic challenges further increase the difficulty of breast lesion segmentation in ultrasound images. To further evaluate the robustness of the proposed method, we conduct experiments on manually degraded BUSI and Dataset B datasets. Specifically, we first add multiplicative noise with $\sigma=0.2$ to the original ultrasound images. Subsequently, we use a blurring kernel with kernel size is $5\times5$ to blur the ultrasound image. The final degraded ultrasound images are shown in Fig. 9. The quantitative evaluation results of different methods on degraded ultrasound images are shown in Table 7. From Table 7, it can be seen that our method achieved the most satisfactory results on two degraded datasets. Although SKU-net has higher specificity values on the degraded BUSI dataset, the results of the remaining metrics are not ideal. According to Table 4 and Table 7, it can be seen that the results of the proposed method in this paper exhibit very little fluctuation on the original ultrasound image and the degraded ultrasound image. This shows that our method has strong capability to cope with noise and blurring. We can also see from Fig. 9 that our method has fewer missed and false detections.

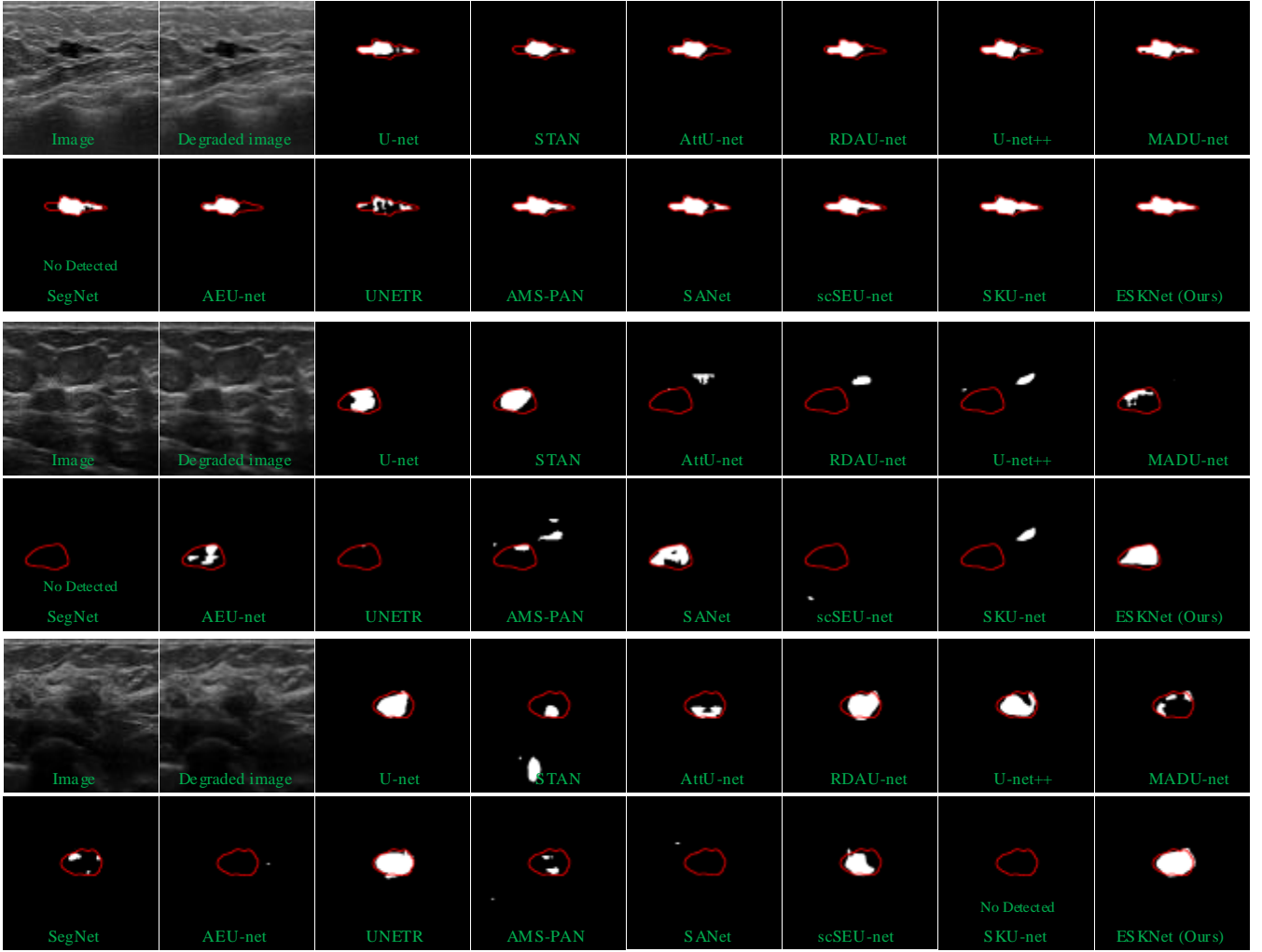

Fig. 9. The prediction masks on the manually degraded BUSI and Dataset B datasets (with multiplicative noise $\sigma=0.2$ and blur kernel size is $5\times5$).

### 6.1.4 Kidney ultrasound application

To further evaluate the robustness of ESKNet, which is used to kidney ultrasound image segmentation task. The kidney ultrasound dataset includes 300 images of the patient with kidney cyst from the Fourth Medical Center of the PLA General Hospital and the Civil Aviation General Hospital. These images are collected by Esaote MyLab, Hitachi, and Philips EPIQ7 ultrasound systems. The case images acquired by single system are 100. The kidney and cyst masks in these ultrasound images were obtained by manual annotation from two urologists. These two doctors have extensive clinical experience and specialize in the interpretation of ultrasound images. Similarly, these ultrasound images were resized to have the same size $512\times512$ for training and testing. We employ three-fold cross-validation to train our method and comparative methods. Similar to breast ultrasound training, we performed data augmentation operations for each fold of training data. In the comparison experiments, the segmentation results of all methods are not subjected to any post-processing.

Table 8 shows the scores of various segmentation methods on the evaluation metrics for kidney ultrasound image segmentation. Our method achieved the highest scores on five evaluation metrics. In general, our method achieves the most satisfying segmentation performance on the kidney ultrasound image. Fig. 10 demonstrates the segmentation results of different methods on renal ultrasound images. It can be seen from Fig. 10 that our method can effectively alleviate the disturbance of uneven energy distribution, blurred boundaries, and similar surrounding tissue to the segmentation results.

**Table 8** The quantitative evaluation results (mean ± std) on kidney ultrasound dataset. The top 2 scores are marked with bold and blue text, respectively. Asterisks means that the P-value of t-test is less than 0.05.

| Methods | U-net | STAN | RDAU-Net | Att U-net | UNETR | AMS-PAN | U-net++ | MADU-net | SegNet | AE U-net | SKU-net | Ours |
|---|---|---|---|---|---|---|---|---|---|---|---|---|
| Jaccard | 71.34±0.88 | 75.89±1.0.2 | 76.87±0.91 | 72.15±1.71 | 51.60±4.63 | 72.13±1.60 | 74.93±2.96 | 73.68±3.03 | 79.94±1.52 | 79.43±1.38 | 80.51±1.33* | **82.19±0.63** |
| Precision | 82.58±1.00 | 84.03±1.55 | 86.08±0.57 | 80.40±1.76 | 66.90±7.03 | 80.71±2.49 | 81.20±2.86 | 82.13±2.83 | 87.19±1.99 | 86.37±1.77 | 88.41±0.34* | **89.66±0.72** |
| Recall | 84.58±0.96 | 89.31±0.31 | 87.92±0.85 | 88.58±0.81 | 72.43±7.01 | 87.88±1.03 | **91.45±0.67** | 88.57±1.66 | 90.89±0.34 | 91.08±0.73 | 90.56±1.49 | 90.99±0.76 |
| Specificity | 97.75±0.05 | 97.98±0.15 | 98.15±0.09 | 97.48±0.20 | 95.74±0.71 | 97.47±0.21 | 97.63±0.33 | 97.70±0.29 | 98.40±0.11 | 98.25±0.14 | 98.45±0.09* | **98.65±0.07** |
| Dice | 82.03±0.70 | 85.01±1.07 | 85.79±0.66 | 82.52±1.44 | 66.49±4.12 | 82.62±1.26 | 84.01±2.35 | 83.57±2.20 | 87.32±1.56 | 87.39±1.34 | 88.41±0.92* | **89.31±0.58** |

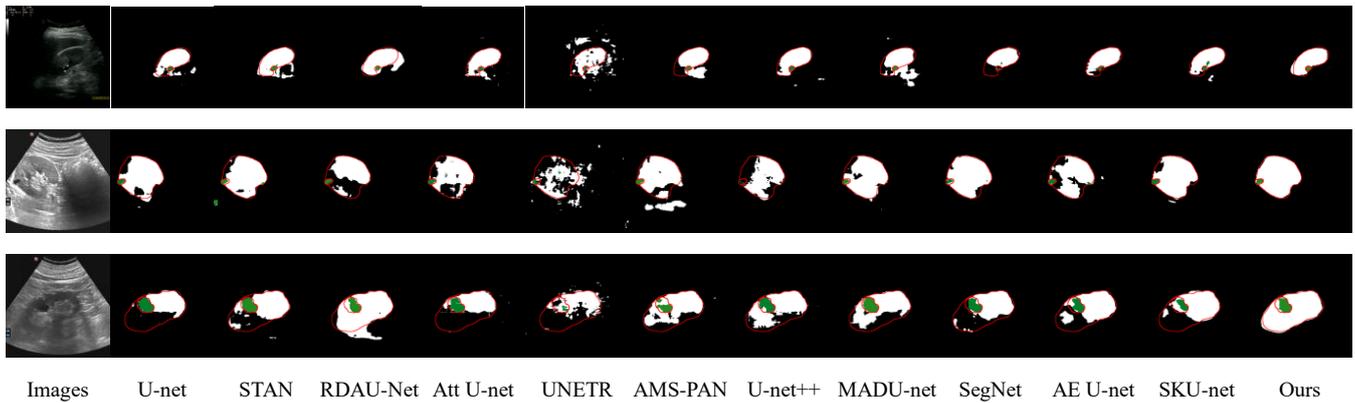

Fig. 10. The visualization of kidney ultrasound images prediction masks by different approaches.

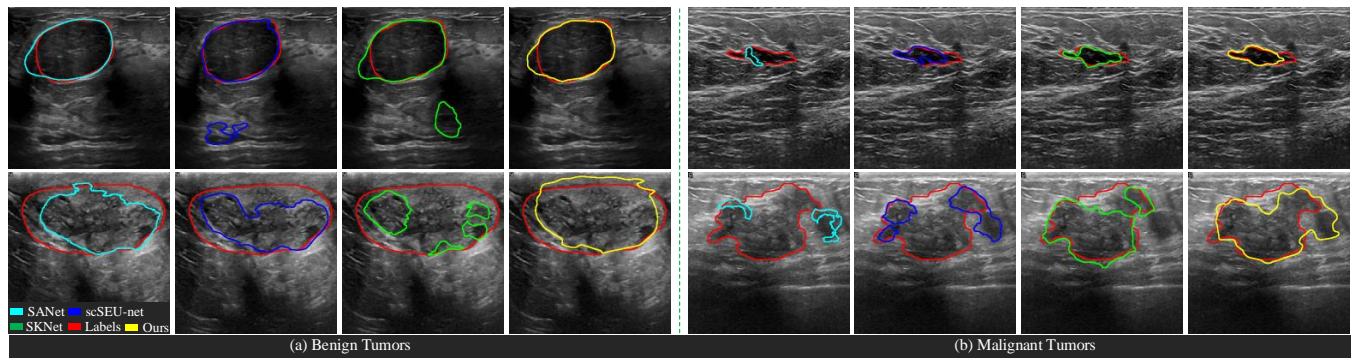

Fig. 11. Some breast lesions that are not well segmented by our approach. In addition, the prediction results of the three methods that are competitive in the comparison methods are displayed. The red lines represent the ground-truth masks.

### 6.2 Limitations and feature works

A large number of experiments have demonstrated the advancement of the study in breast tumor segmentation, but there are still some limitations that need further research: (i) Although the design of ESKNet has further improved the segmentation accuracy of breast tumors, it still does not completely overcome the limitations of the complexity of ultrasound pattern on clinical practice. As shown in Fig. 11, Our method also has some missing and false detection in the segmentation of breast lesions. (ii) The spatial and channel attention mechanism of the enhanced selective kernel convolution (ESK) module improves the representation ability of breast tumors, but inevitably increases the complexity of the network compared with the original selective kernel convolution (SK) module, as shown in Table 9. To alleviate the above challenges, we still need to put efforts in the following two directions in future research: (i) We can cope with the challenge of missed or false detection by refining residual learning strategies. Specifically, we can first introduce boundary loss and region loss in the backbone network to obtain a rough map closer to the ground-truth mask. Then, develop a missed detection residual network and a false detection residual network to capture the missed residual maps and false residual maps, respectively. Finally, the rough map captured by the backbone network is integrated with the missed residual map and false residual map to obtain the final refined prediction mask. (ii) It is necessary to further reduce the complexity of modules while ensuring the robustness of the network. Specifically, we can first reduce the number of ESK modules used. Then, we can use the salient maps of breast ultrasound images as inputs into the network to ensure the performance of the network. The salient map of breast ultrasound can be referred to SMU-net (Ning et al., 2021).

Table 9 Complexity analysis of different segmentation methods in terms of network parameters and computational costs.

| | U-net | STAN | Att U-net | RDAU-Net | UNETR | AMS-PAN | U-net++ | MADU-net | SegNet | AE U-net | SANet | scSEU-net | SKU-net | Ours |
|---|---|---|---|---|---|---|---|---|---|---|---|---|---|---|
| Params (M) | 7.85 | 29.80 | 35.56 | 51.76 | 112.11 | 26.40 | 36.63 | 39.85 | 42.02 | 19.73 | 16.47 | 13.42 | 34.02 | 44.57 |
| GFLOPs | 62.82 | 254.18 | 141.45 | 133.13 | 295.21 | 768.81 | 620.44 | 218.32 | 228.23 | 240.95 | 118.02 | 128.18 | 250.82 | 491.94 |

## 7. Conclusion

This paper develops a novel enhanced selective kernel convolution (ESK) based on selective kernel convolution (SK) to segment breast lesions. The ablation experiments on the BUSI and Dataset B datasets fully demonstrate the effectiveness of each network component in the task of breast lesion segmentation. The visualization results also show that the introduction of different components can help the network to obtain a predictive mask that is closer to the ground-truth label. Comparison with fourteen state-of-the-art deep learning segmentation methods further illustrates the superiority of the proposed approach. In addition, we also fully demonstrate the generalization ability of the proposed method in this paper through "Robustness on Benign and Malignant Lesions", "External validation", "Robustness on degraded ultrasound images" and "Kidney ultrasound application" experiments. Although our proposed method suffers from a certain amount of missed detection and false detection on individual images, it still achieves the most satisfactory results. In summary, the adequate experimental results fully demonstrate the superior performance of our method for segmenting breast ultrasound images.

**Acknowledgments**

This work is supported by the National Natural Science Foundation of China (U1913207) and the Tianjin Research Innovation Project for Postgraduate Students (2022BKY004). Special thanks to Dr. Lei Li, University of Oxford, and Professor MoiHoon Yap, Manchester Metropolitan University, for supporting the work on this paper.